\documentclass[aps,prx,reprint,twocolum,amsmath,superscriptaddress,amssymb,showpacs,nofootinbib]{revtex4-2}

\usepackage[utf8]{inputenc}
\usepackage[english]{babel}
\usepackage{booktabs} 
\usepackage{comment} 
\usepackage{hyperref} 
\usepackage{physics} 
\usepackage[table,xcdraw,dvipsnames]{xcolor} 

\allowdisplaybreaks

\usepackage{lipsum}
\usepackage{xfrac}

 \usepackage{multirow}

\usepackage{amsmath,verbatim,latexsym,amssymb,indentfirst,mathrsfs,mathtools,amsthm,bbm,bm,hyperref,url,cancel,subcaption}
\usepackage[font=small,labelfont=bf,format=plain,justification=raggedright,singlelinecheck=false]{caption}
\usepackage{verbatim,indentfirst}
\usepackage[title,titletoc]{appendix}

\usepackage{graphicx}
\graphicspath{{images/}}
\usepackage{epstopdf}
\newcommand{\caphead}[1]{{\bf #1}}

\renewcommand{\thesection}{\Roman{section}}
\renewcommand{\thesubsection}{\Roman{section} \Alph{subsection}}
\renewcommand{\thesubsubsection}{\Roman{section} \Alph{subsection} \arabic{subsubsection}}
\makeatletter
\def\p@subsection{}
\makeatother
\makeatletter
\def\p@subsubsection{}
\makeatother

\makeatletter
\newcommand\footnoteref[1]{\protected@xdef\@thefnmark{\ref{#1}}\@footnotemark}
\makeatother

\usepackage{soul}

\usepackage{color}
\usepackage[dvipsnames]{xcolor}
\usepackage[normalem]{ulem}

\newcommand{\LParen}{ \bm{(} }
\newcommand{\RParen}{ \bm{)} }

\renewcommand\th{ {\rm th} }

\makeatletter
\DeclareFontFamily{OMX}{MnSymbolE}{}
\DeclareSymbolFont{MnLargeSymbols}{OMX}{MnSymbolE}{m}{n}
\SetSymbolFont{MnLargeSymbols}{bold}{OMX}{MnSymbolE}{b}{n}
\DeclareFontShape{OMX}{MnSymbolE}{m}{n}{
    <-6>  MnSymbolE5
   <6-7>  MnSymbolE6
   <7-8>  MnSymbolE7
   <8-9>  MnSymbolE8
   <9-10> MnSymbolE9
  <10-12> MnSymbolE10
  <12->   MnSymbolE12
}{}
\DeclareFontShape{OMX}{MnSymbolE}{b}{n}{
    <-6>  MnSymbolE-Bold5
   <6-7>  MnSymbolE-Bold6
   <7-8>  MnSymbolE-Bold7
   <8-9>  MnSymbolE-Bold8
   <9-10> MnSymbolE-Bold9
  <10-12> MnSymbolE-Bold10
  <12->   MnSymbolE-Bold12
}{}

\let\llangle\@undefined
\let\rrangle\@undefined
\DeclareMathDelimiter{\llangle}{\mathopen}%
                     {MnLargeSymbols}{'164}{MnLargeSymbols}{'164}
\DeclareMathDelimiter{\rrangle}{\mathclose}%
                     {MnLargeSymbols}{'171}{MnLargeSymbols}{'171}
\makeatother
\usepackage{amsmath}
\usepackage[normalem]{ulem}
\usepackage{afterpage}
\usepackage{xcolor}[dvipsnames]

\newcommand{\superket}[1]{|#1\rrangle}
\newcommand{\superbra}[1]{\llangle #1 |}
\newcommand{\sw}{\mathrm{Sw}}

\usepackage{soul}

\begin{document}

\title{Critical phase and spin sharpening in SU(2)-symmetric monitored quantum circuits}

\author{Shayan Majidy} 
\email{smajidy@uwaterloo.ca}
\affiliation{Perimeter Institute for Theoretical Physics, Waterloo, Ontario N2L 2Y5, Canada} 
\affiliation{Institute for Quantum Computing, University of Waterloo, Waterloo, Ontario N2L 3G1, Canada}

\author{Utkarsh Agrawal}
\email{utkarsh_agrawal@ucsb.edu}
\affiliation{Kavli Institute for Theoretical Physics, University of California, Santa Barbara, CA 93106, USA}

\author{Sarang Gopalakrishnan}
\affiliation{Department of Electrical and Computer Engineering, Princeton University, Princeton, NJ 08544, USA}

\author{Andrew C. Potter}
\affiliation{Department of Physics and Astronomy, and Quantum Matter Institute, University of British Columbia, Vancouver, BC, Canada V6T 1Z1}

\author{Romain Vasseur}
\affiliation{Department of Physics, University of Massachusetts, Amherst, MA 01003, USA}

\author{Nicole Yunger Halpern}
\email{nicoleyh@umd.edu}
\affiliation{Joint Center for Quantum Information and Computer Science, NIST and University of Maryland, College Park, MD 20742, USA}
\affiliation{Institute for Physical Science and Technology, University of Maryland, College Park, MD 20742, USA}

\date{\today}

\begin{abstract} 
Monitored quantum circuits exhibit entanglement transitions at certain measurement rates. Such a transition separates phases characterized by how much information an observer can learn from the measurement outcomes. We study SU(2)-symmetric monitored quantum circuits, using exact numerics and a mapping onto an effective statistical-mechanics model. Due to the symmetry's non-Abelian nature, measuring qubit pairs allows for nontrivial entanglement scaling even in the measurement-only limit. We find a transition between a volume-law entangled phase and a critical phase whose diffusive purification dynamics emerge from the non-Abelian symmetry. Additionally, we identify a ``spin-sharpening transition.'' Across the transition, the rate at which measurements reveal information about the total spin quantum number changes parametrically with system size.
\end{abstract}

\maketitle

\section{Introduction}\label{sec:int}

Traditionally, quantum many-body physicists have been limited to studying closed systems in equilibrium. Thanks to the maturation of quantum simulators~\citep{preskill2018quantum}, researchers can now prepare and control open quantum systems far from equilibrium with high precision. Quantum simulators have helped answer foundational questions about quantum entanglement and thermodynamics \citep{abanin2019colloquium, altman2021quantum}. Also, quantum simulators have the potential to solve real-world problems in, e.g., materials science and chemistry~\citep{daley2022practical, altman2021quantum}. These advances have {raised questions about open quantum systems and, in turn, the role of measurements in quantum dynamics}~\cite {schaller2014open}.{ Some of these questions have been answered using \textit{monitored quantum circuits}}~\cite{fisher2022random, potter2022entanglement}, {which combine many-body dynamics and measurements.} A typical monitored quantum circuit acts on a chain of $L$ qubits (spin-$1/2$ particles). The circuit contains two-qubit unitary gates, after each layer of which every qubit has a probability $p$ of being measured. Monitored circuits exhibit measurement-induced phase transitions (MIPTs), due to the competition between chaotic dynamics and measurements ~\citep{gullans2020dynamical,gullans2020scalable,jian2020measurement,bao2020theory,szyniszewski2019entanglement,zabalo2020critical,zabalo2022operator,lopez2020mean,weinstein2022measurement,li2021conformal,turkeshi2020measurement, khemani2018operator, rakovszky2018diffusive, agrawal2022entanglement, sang2021measurement, li2021robust, lavasani2021measurement, ippoliti2021entanglement, lavasani2021topological, chan2019unitary,cao2019entanglement,nahum2021measurement, chen2020emergent, kelly2022coherence}. 

Initially, an MIPT was cast as a transition between phases characterized by volume-law and area-law entanglement scaling~\citep{skinner2019measurement, li2019measurement}. Equivalently, the transition is a purification transition between a \textit{mixed phase} and a \textit{pure phase}~\citep{gullans2020dynamical}. When the measurement rate $p$ is low, the chaotic dynamics scramble information about the initial state. Local measurements cannot extract that information in this mixed phase. An initially mixed state becomes pure, conditionally on measurement outcomes, in a time $t_{\rm P} \sim \exp(L)$, with $L$ the number of qubits. In contrast, at large $p$, the measurements can distinguish different initial states efficiently. In this pure phase, an initially mixed state purifies quickly, often at an $L$-independent rate~\citep{skinner2019measurement}.

Few properties restrict the simplest monitored circuits' dynamics: unitarity and locality. Monitored circuits can be enriched, though. Enhancements include charge conservation \cite{khemani2018operator, rakovszky2018diffusive, agrawal2022entanglement, sang2021measurement, li2021robust, lavasani2021measurement}, measurements of particular operators (such as generators of the toric-code stabilizer) \cite{ippoliti2021entanglement, lavasani2021topological}, and the replacement of qubits with free fermions~\cite{chan2019unitary,cao2019entanglement,nahum2021measurement, chen2020emergent, alberton2021entanglement, nahum2020entanglement}. U(1)-symmetric monitored circuits exhibit a \textit{charge-sharpening transition} \citep{agrawal2022entanglement} between a \textit{charge-fuzzy phase} and a \textit{charge-sharp phase}. These phases are distinguished by how quickly measurements collapse superpositions of different amounts of charge---how efficiently an observer can learn from local  measurements the amount of global charge in the system.

Noncommuting symmetry charges have spawned a growing subfield of quantum thermodynamics~\citep{majidy2023noncommuting, lostaglio2017thermodynamic,guryanova2016thermodynamics,nyh2016microcanonical}. Noncommutation of charges has been shown to increase average entanglement~\cite{majidy2023non}, decrease entropy-production rates~\cite{manzano2018Squeezed,Upadhyaya_23_What}, and necessitate modifications to the eigenstate thermalization hypothesis (ETH)~\cite{srednicki1994chaos, murthy2022non}. Researchers have used trapped ions to bridge this subfield from theory to experimental reality~\cite{kranzl2022experimental,nyh2020noncommuting,nyh2022how}. This subfield's discoveries partially motivates our work, as do two computational results: first, a model of quantum computation can be defined from SU(2)-symmetric gates and spin fusions~\cite{jordan2009permutational}. Second, SU(2)-symmetric measurements can achieve universal quantum computation, if performed on certain initial states~\citep{rudolph2005relational, rudolph2021relational}. 
We therefore study monitored quantum circuits with three noncommuting charges. 

In this work, we explore monitored-random-circuit dynamics of one-dimensional (1D) qubit chains with SU(2) symmetry. Equivalently, the circuits conserve three noncommuting charges: the total spin angular momentum's components. First, we explore the purification dynamics of a spin chain initially entangled with an ancilla spin. We identify a purification transition between a mixed phase, in which the ancilla purifies over an exponential-in-$L$ time, and a critical phase\footnote{For the purposes of this discussion, we classify Goldstone phases with long-range order as critical.} with scale-invariant purification and entanglement growth. Above a critical measurement rate (at $p>p_{\rm c}$), we observe an extended-in-$p$ critical phase in which the purification time scales diffusively: $t_{\rm P}\sim L^2$. Second, we examine the entanglement dynamics undergone by an initially unentangled state. The purification transition doubles as an entanglement transition between volume-law entanglement scaling, at $p<p_{\rm c}$, and subextensive (logarithmic or small-power-law) scaling, at $p>p_{\rm c}$. The critical entanglement dynamics $p>p_{\rm c}$---even in the measurement-only limit ($p=1$)---due to the local measurements' noncommuting nature. In fact, a Lieb-Shultz-Mattis-type anomaly precludes a simple area-law entangled regime~\cite{nahum2020entanglement}, as would arise when $p=1$, absent symmetries.

Observing the purification/entanglement transition experimentally would require many instances of the same set of measurement outcomes. Such instances occur with vanishing likelihood in the thermodynamic limit. This challenge is the \textit{postselection problem}. To evade this difficulty, we explore a ``spin-sharpening/learnability" transition. Denote by $s$ the total spin quantum number. We examine whether the dynamics collapse an initial superposition of states in different $s$ sectors. Unlike in the U(1)-symmetric problem, the sectors generally cannot be shared by the (extensive) charges: our system’s three charges, failing to commute, share only one sector. We identify a spin-sharpening transition at a measurement rate $p=p_\#$, which is numerically indistinguishable from the entanglement-transition rate: $p_\#\approx p_{\rm c}$.
In the ``spin-sharp'' phase ($p>p_\#$), an observer can, in principle, determine the system's $s$ in a time scale $t\sim L^2$, with a probability tending to unity as $L\rightarrow \infty$. In contrast, in the ``spin-fuzzy'' phase ($p<p_\#$), 
the time scale is $t \sim L^3$. This ``learning'' perspective might be used to probe the transition experimentally.

Finally, we interpret our results within an effective replica statistical-mechanics model. We obtain the model by averaging over the gates and measurement outcomes, building on previous results about asymmetric and symmetric circuits~\citep{potter2022entanglement}. This model casts dynamical properties of SU(2)-symmetric monitored quantum circuits in terms of some effective Hamiltonian's low-energy properties. We interpret our numerical results in terms of this effective-Hamiltonian model.

The rest of this paper is organized as follows. In Sec.~\ref{sec:SU2}, we introduce SU(2)-symmetric monitored quantum circuits. We present the purification/entanglement transition in Sec.~\ref{sec:Num} and the spin-sharpening transition in Sec.~\ref{Sec:ChaSha}. Section~\ref{Sec:StaMecMod} contains our statistical-mechanics mapping. Section~\ref{sec:OutLoo} finishes with opportunities established by this work.

\section{Model: SU(2)-symmetric monitored circuits}
\label{sec:SU2}

\begin{figure}
    \centering
    \includegraphics[width=0.8\columnwidth]{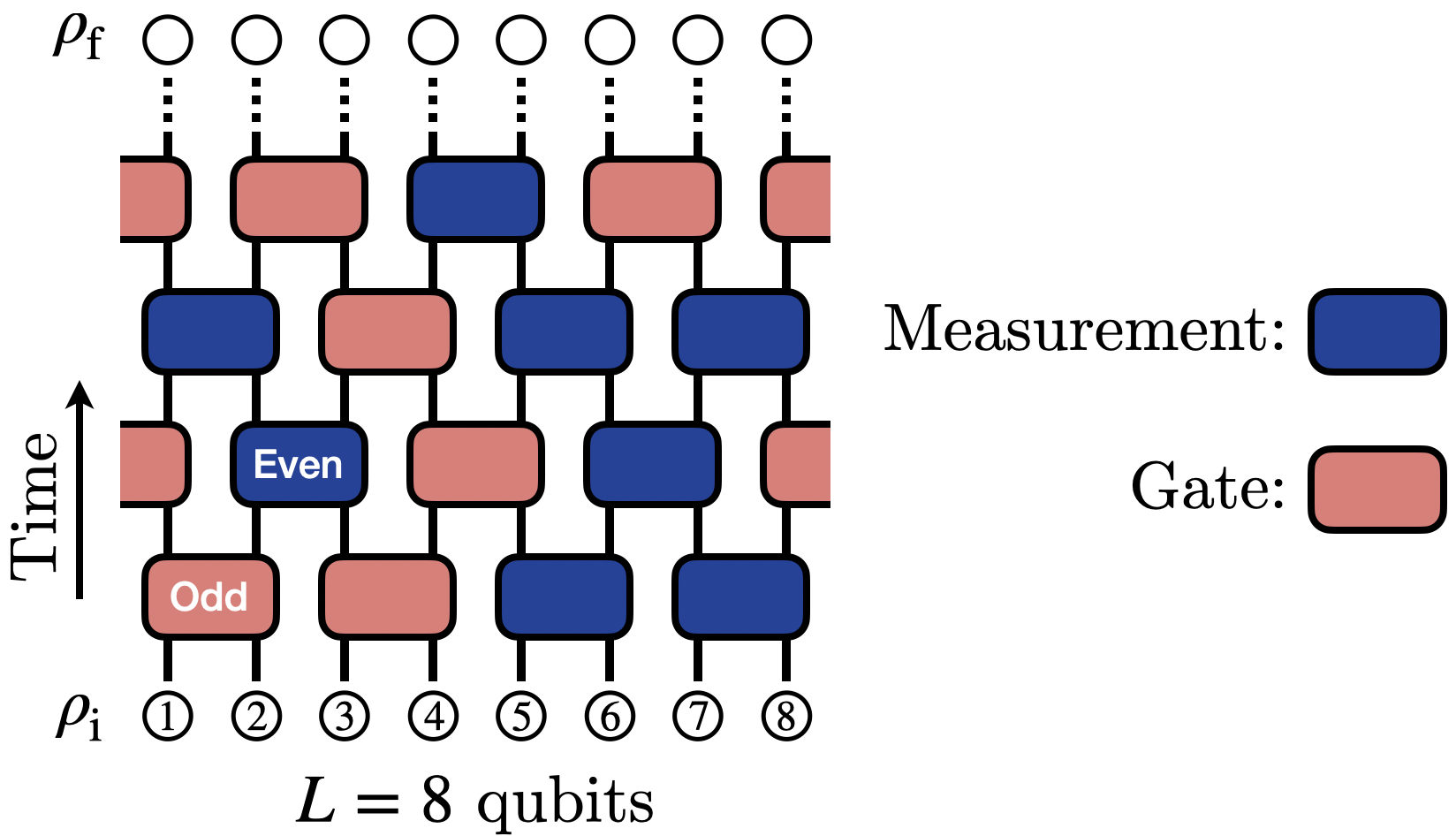}
    \caption{\caphead{SU(2)-symmetric monitored quantum circuits.} 
    $L$ qubits (circles) are prepared in the state $\rho_{\rm i}$. Each ``brick'' in the brickwork circuit is an SU(2)-symmetric unitary gate with a probability $1-p$ and is a two-qubit projective measurement with a probability $p$. The circuit acts for some time (some number of layers) before the final state, $\rho_{\rm f}$, is read out. One brick illustrates which bonds have even (odd) indices.
    }
    \label{fig:circuit_models}
\end{figure}

Consider a brickwork circuit acting on a 1D chain of qubits, as depicted in  Fig.~\ref{fig:circuit_models}. The number $L$ of spins is even for convenience.
Denote by $\sigma_j^{(x,y,z)}$ the Pauli matrices acting on qubit $j$.
The total spin components
$S^{(x,y,z)} = \frac{1}{2} \sum_{j=1}^L \sigma_j^{(x,y,z)}$ generate the algebra associated with a global SU(2) symmetry. We set $\hbar$ to 1. The spin-squared operator $\vec{S}^2$ has eigenvalues $s(s+1)$ labelled by the total spin quantum number $s$. We denote the eigenvalues of $S^{(z)}$ by $m$, the two-qubit singlet state by $\ket{ {\rm s}_0}$, and the two-qubit eigenvalue-$m$ triplets by $\ket{ {\rm t}_m}$.

Each brick is, with a probability $1-p$, a gate,  or, with a probability $p$, a projective measurement. The gates are chosen randomly from SU(2). The most general such gate acting on spins $j$ and $j+1$, has the form
\begin{align}
    \cos(\phi)\mathbb{I}-i\sin (\phi) ~\sw_{j,j+1}, \label{eq:SU2unitary}
\end{align}
up to an irrelevant overall phase.
$\sw_{j,k}$ swaps the states of the spins $j$ and $k$. We draw each gate's parameter $\phi$ independently from the uniform distribution on $[0,2\pi)$. 
Each measurement projects a two-qubit state onto the singlet ($s=0$) or triplet ($s=1$) subspace (fusion channel). Crucially, two measurements fail to commute when acting on overlapping spin pairs. {Thus,} the SU(2) symmetry precludes nontrivial single-qubit measurements. One time step consists of a brick layer on even-index bonds and a layer on odd-index bonds. In the even-index-bond layers, a brick connects the first and $L^{\th}$ qubits, effecting periodic boundary conditions.

\section{Purification/entanglement transition} \label{sec:Num}

We first explore the model's entanglement and purification dynamics. In Sec.~\ref{sec:PurTim}, we examine the purification dynamics of initially mixed states. The spin chain's state begins scrambled and entangled with an ancilla qubit $A$. The ancilla's entanglement decreases over time. In Sec.~\ref{sec:EntDyn}, we examine the entanglement dynamics exhibited by initially short-range-entangled pure states.

\subsection{Purification time}\label{sec:PurTim}

\begin{figure*}
    \centering
    \includegraphics[width=2\columnwidth]{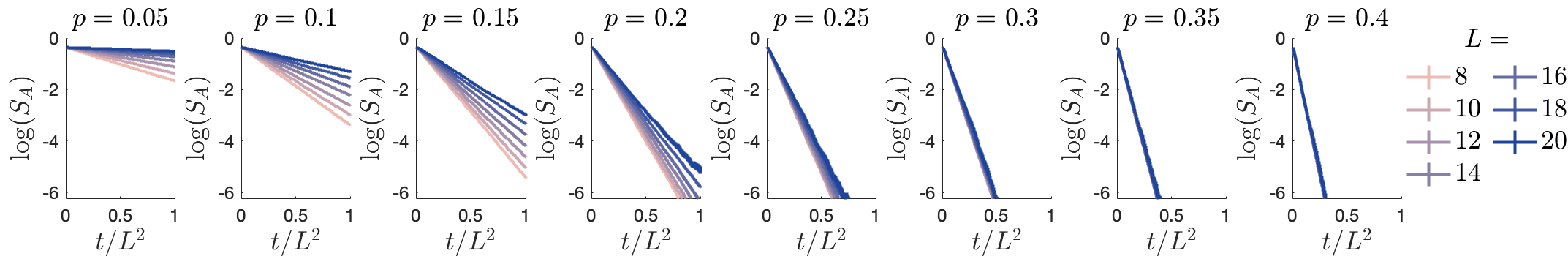}
    \caption{\caphead{The purification time reveals a $z{=}2$ phase.} The entropy $S_A$ quantifies the ancilla qubit's entanglement with the system. We plot $\log(S_A)$ for clarity, as $S_A$ decays exponentially. $t/L^2$ runs along the $x$-axis to demonstrate the existence of a phase in which the system purifies over a time scale $t_{\rm P} \sim L^2$. The curves' collapsing at $p> 0.35$ evidences this phase. We used 30\,000 samples when $L = 8$ to $16$; 10\,000 samples when $L = 18$; and 1\,500 samples when $L=20$. The $y$-axis's lower limit is $\log(10^{-3}) \approx -6.91$. {Additional numerics for $p=0.6, 0.8$, and $1.0$ are included in Appendix} \ref{sec:AddNum}.
    } 
    \label{fig:purification_times}
\end{figure*}

We determine the purification time as follows~\citep{gullans2020scalable}. Denote by $\ket{s,m,\lambda_0}$ and $\ket{s,m,\lambda_1}$ two orthogonal states from the same $(s,m)$ sector. The last index distinguishes degenerate states. We entangle an ancilla qubit with the system's $L$ qubits, forming the $(L+1)$-qubit state
\begin{equation}
   \label{eq:PurTimeInitial}
    \tilde{\ket{\psi_{\rm i}}} = \tfrac{1}{\sqrt{2}} \left(  \ket{0}_{A}\ket{s,m,\lambda_0} + \ket{1}_{ A}\ket{s,m,\lambda_1}\right). 
\end{equation}
The subscript $A$ distinguishes the ancilla from the system qubits. $A$ does not undergo gates or measurements. 

We choose two system states that have $s=1$ and $m=0$. In  $\ket{{s=1}, {m=0}, \lambda_0}$, qubits $1$ and $2$ are in the triplet $\ket{ {\rm t}_0}$; and the remaining pairs of qubits, in singlets $\ket{ {\rm s}_0}$. In $\ket{s=1,m=0,\lambda_1}$, qubits $3$ and $4$ are in $\ket{ {\rm t}_0}$, instead. These two system states are orthogonal, in the same $\vec{S}^2$ sector, and in the same $S^{(z)}$ sector. However, one can distinguish the states by measuring qubits $1$ and $2$. Such local distinguishability is undesirable. Therefore, after preparing $\tilde{\ket{\psi_{\rm i}}}$, we scramble the system: the system undergoes a unitary-only ($p=0$) SU(2)-symmetric circuit for $L^2$ time steps. 
(The $t_{\rm P}$ identified later in this subsection
motivates the $L^2$.) The scrambling encodes quantum information about the ancilla roughly uniformly in many-body entanglement. This process prepares $\ket{\psi_{\rm i}}$. 

$\ket{\psi_{\rm i}}$ undergoes $t=L^2$ time steps under monitored-random-circuit dynamics with $p \geq 0$. Denote by $\rho_{A} \coloneqq \Tr_{\bar{A}}(\dyad{\psi_{\rm f}})$ the final state of $A$. We calculate the final entanglement entropy between $A$ and the system:
\begin{equation}
    S_A \coloneqq S(\rho_{A}) \coloneqq - \Tr[\rho_{A} \log(\rho_{A})]. \label{eq:EE_onestate}
\end{equation}
(All logarithms are base-$e$.) We anticipate that the measurements will purify the system at an exponential-in-$t$ rate: $S_A\sim e^{-t/t_{\rm P}(L)}$. Therefore, we plot $\log (S_A)$ in Fig.~\ref{fig:purification_times}. Along the $x$-axes runs $t/L^2$. At each $p > p_{\rm c} \approx 0.35$, 
the different-$L$ curves collapse.
Hence this phase purifies according to $S_A \sim {\rm e}^{-  t/L^2}$ and so has a dynamical critical exponent $z{=}2$. This $z$-value characterizes diffusive scaling~\citep{cordery1981physics} and suggestively evokes ferromagnetic spin waves' dynamics~\citep[Ch.~33]{ashcroft2022solid}.

\begin{figure}
    \centering
    \includegraphics[width=0.67\columnwidth]{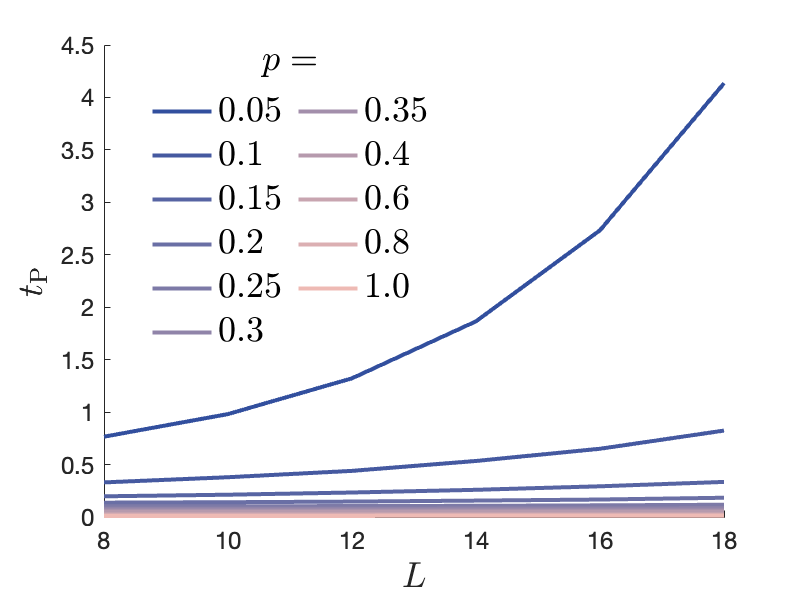}
    \caption{\caphead{{Qualitative comparison of the purification time's growth with $L$ at different $p$ values.}} {For $p<p_c$, the purification time diverges rapidly with system size in a manner consistent with exponential.}}
    \label{fig:purification}
\end{figure}

At lower measurement rates, $p \ll p_{\rm c}$, we observe a mixed phase. 
Figure~\ref{fig:purification} shows the purification time plotted against $L$, at several $p$ values. At $p = 0.05$, $t_{\rm P} \sim e^{L}$. At $p$ values between 0.05 and 0.35, the scaling is unclear from the numerics; we cannot conclude to which phase this intermediate regime belongs. Still, the exponential purification time resembles that of asymmetric circuits~\citep{gullans2020dynamical}.
We analyze this mixed phase analytically in Sec.~\ref{Sec:StaMecMod}, using a duality between monitored circuits and a statistical-mechanics model.

\subsection{Entanglement dynamics}\label{sec:EntDyn}

To characterize the critical phase further, we explore an initially pure state's late-time bipartite entanglement entropy, $S_{\rm f}$. The purification transition manifests as a qualitative change in the $L$-dependence of $S_{\rm f}$ at $p=p_{\rm c}$.

\begin{figure}
    \centering    \includegraphics[width=0.66\columnwidth]{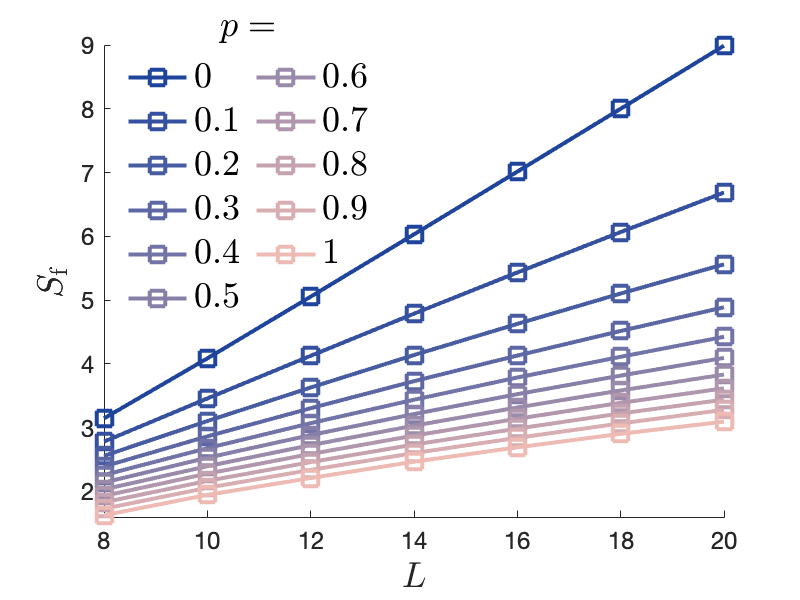}
    \caption{\caphead{The entanglement dynamics evidence no area-law phase.}
    The bipartite entanglement entropy reaches the long-time value $S_{\rm f}$. At $p=0$, $S_{\rm f}$ is linear in $L$. As $p$ increases, $S_{\rm f}$ gradually becomes logarithmic or power-law with a small exponent. When $L = 10$ to $16$, we use 30\,000 samples; when $L = 18$ and $20$, we use 10\,000.
    }    \label{fig:Ss_v_L_diff_p}
\end{figure}

We initialize the system in a short-range-entangled state
$\ket{\psi_{\rm i}}$, a tensor product of a triplet $\ket{ {\rm t}_0}$ and $\tfrac{L-2}{2}$ singlets $\ket{ {\rm s}_0}$. This choice's details are unimportant. However, we choose this state so that $\ket{\psi_{\rm i}}$ is in the same $s$ sector at all system sizes $L$. The state undergoes monitored-circuit dynamics for $L^2$ time steps. This time suffices for the entanglement entropy to reach a steady value, regardless of the measurement rate, $p$.
Figure~\ref{fig:EE_vs_time} in Appendix \ref{sec:AddNum} illustrates this point at the extreme values $p=0,1$.
We measure the bipartite entanglement entropy, $S_{\rm f}$, between two equal-size halves of the chain.

Figure~\ref{fig:Ss_v_L_diff_p} shows the dependence of $S_{\rm f}$ on $L$ at different measurement rates $p$. At $p=0$, we observe the volume-law phase common to monitored circuits: $S_{\rm f} \sim L$. Figure~\ref{fig:example_fits} in Appendix~\ref{sec:AddNum} supports this claim more precisely than does Fig.~\ref{fig:Ss_v_L_diff_p}. At larger $p$ values,\footnote{According to~\cite{gullans2020dynamical}, the entanglement phase transition is equivalent to the purification transition. Our system's purification transition happens at $p_{\rm c} \approx 0.35$, according to the previous subsection. This section's numerics are consistent with an entanglement transition at $p \approx 0.35$, but the transition's exact location is unclear.}
the entanglement scaling is less consistent with a linear fit. Better fits are logarithmic and small-power-law $(S_{\rm f} \sim \sqrt{L})$. One cannot definitively distinguish these behaviors at the accessible system sizes, as detailed in Appendix~\ref{sec:AddNum}.
The statistical-mechanics model (Sec.~\ref{Sec:StaMecMod}) provides stronger evidence for the absence of a volume-law phase at large $p$.


Intuitively, the slow entanglement growth at large $p$, even in the measurement-only ($p=1$) limit, arises from the noncommutativity of the charges measured. 
Similar logarithmic entanglement growth has been observed under measurement-only dynamics previously: Majorana fermions were subjected to noncommuting measurements in~\cite{ippoliti2021entanglement}.


Appendix~\ref{sec:Cor} presents numerical results concerning the correlations between local observables at different sites. The limitation on system size makes it difficult to determine the functional form of the correlations' decay with distance. However, we find a qualitative change in how the correlations decay at $p>p_{\rm c}$ and at $p<p_{\rm c}$.

\section{Spin-sharpening transition}\label{Sec:ChaSha}

Having explored the purification dynamics within an $s$ sector, we explore the purification of a superposition spread across $s$ sectors. We again entangle the chain with an ancilla qubit. This time, the ancilla is in $\ket{0}$, and the chain has a spin quantum number $s_0$, in superposition with the ancilla's being in $\ket{1}$ and the chain's having $s_1$. The dynamics may purify the ancilla in a given measurement trajectory. In this case, the chain's state has collapsed onto the $s_0$ (or $s_1$) sector. Consequently, the measurement outcomes' probability of being compatible with the system's having $s_1$ (or $s_0$) vanishes. An observer 
{with knowledge of the circuit can} learn the spin quantum number by monitoring measurement outcomes {(though doing so may require the ability to classically simulate the circuit with post-selected measurement outcomes).}

Comparing spin sharpening with U(1)-charge sharpening is illuminating. One can estimate as follows the total charge of qubits undergoing a U(1)-symmetric hybrid circuit: Running the circuit, one obtains $ptL$ measurement outcomes (0s and 1s), on average. Consider averaging the outcomes, multiplying by $L$, and rounding to the nearest integer. If $t \sim L$, this procedure estimates the charge accurately~\citep{barratt2022transitions}. If the dynamics are SU(2)-symmetric on average (as in Sec.~\ref{sec:SU2}), sequential measurements fail to commute. Hence later measurements render partially irrelevant the information obtained from earlier measurements. An observer cannot obviously learn $s$ ever. Nevertheless, we numerically identify a measurement-induced transition at a measurement rate $p_\#$.
We call this transition a \emph{spin-sharpening} transition. It separates regimes in which an observer can ($p>p_\#$) and cannot ($p<p_\#$) identify $s$ from the measurement outcomes, with a probability tending to unity as the $L\rightarrow \infty$.

We diagnose the spin-sharpening transition using a similar procedure to the one in Sec.~\ref{sec:PurTim}. The difference is that, unlike in Eq.~\eqref{eq:PurTimeInitial}, we construct $\tilde{\ket{\psi_{\rm i}}}$ from distinct $\vec{S}^2$ eigenspaces:
\begin{equation}
    \tilde{\ket{\psi_{\rm i}}} = \tfrac{1}{\sqrt{2}} \left(  \ket{0}_A\ket{s_0,m,\lambda_0} + \ket{1}_A\ket{s_1,m,\lambda_1}\right).
\end{equation}
We choose $m=0$, $s_0 = 1$, and $s_1 = 0$ for convenience:
one can construct such a $\tilde{\ket{\psi_{\rm i}}}$ by tensoring together singlets and an $m=0$ triplet, regardless of $L$. After preparing $\tilde{\ket{\psi_{\rm i}}}$, we scramble the system under a $p{=}0$ circuit for $L^2$ time steps, as in Sec.~\ref{sec:PurTim}. This procedure prepares a state $\ket{\psi_{\rm i}}$. Then, we evolve the system under monitored-circuit dynamics with a fixed $p$. Anticipating $z{=}2$ dynamical scaling in the spin-sharp phase, we evolve the system for $L^2$ time steps. If the ancilla purifies after this short time, we say that the spin has sharpened. We denote the final state by $\ket{\psi_{\rm f}}$.

\begin{figure}
    \centering
    \includegraphics[width=0.49\columnwidth]{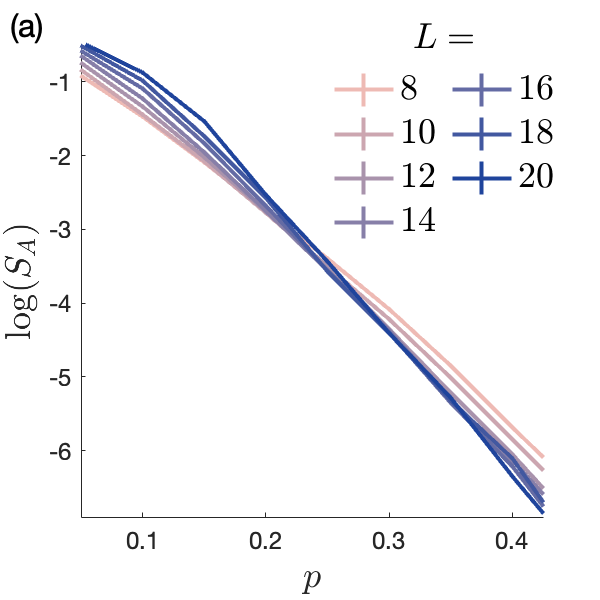}
    \includegraphics[width=0.49\columnwidth]{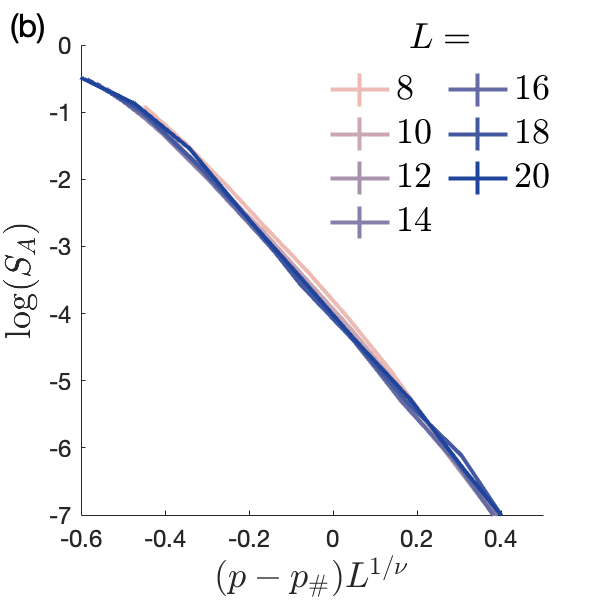}
    \caption{\caphead{{Evidence of} spin-sharpening transition.} The entropy $S_A$ quantifies the ancilla qubit's entanglement with the system. Different curves correspond to different system sizes $L$. (a) The curves' crossing at $p \approx 0.28$ indicates a phase transition. (b) We identify a finite-size collapse using $\nu = 3.0$ and $p_{\#} = 0.28$.
    }
    \label{fig:CS}
\end{figure}

Figure~\ref{fig:CS}a shows the ancilla's final entanglement entropy, $S_A$, plotted against $p$. Different curves correspond to different system sizes $L$. The curves cross at $p_\# \approx 0.28$, suggesting that a spin-sharpening transition occurs at $p_\#$. Furthermore, Fig.~\ref{fig:CS}b displays a finite-size collapse. We used the scaling form $\log(S_A) = (p - p_\#) L^{1/\nu}$, the correlation-length exponent $\nu = 3.0$, and $p_{\#} = 0.28$. {Using $\nu = 3.0$, we observe a suitable collapse. $\nu$ values within $\pm 1.2$ of 3.0 yield reasonable collapses, too.}

\begin{figure}
    \centering
    \includegraphics[width=0.49\columnwidth]{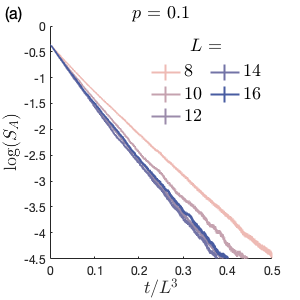}
    \includegraphics[width=0.49\columnwidth]{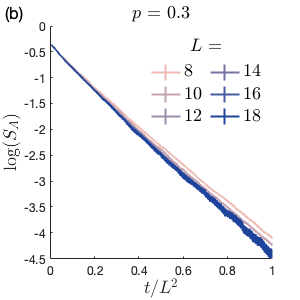}
    \caption{
    \caphead{The spin-sharpening time scale is $\sim L^3$ in the fuzzy phase and $\sim L^2$ in the sharp phase.} The entropy $S_A$ quantifies the ancilla qubit's entanglement with the system. Different curves correspond to different system sizes $L$. (a) $t/L^3$ runs along the $x$-axis to demonstrate that the spin can sharpen over a time scale $\sim L^3$. This time scale characterizes the {spin}-fuzzy phase ($p < p_{\#}$). 
    {Simulating an $L{=}18$ circuit over $L^3$ time steps is not computationally feasible. Thus no $L=18$ curve is present.} (b) $t/L^2$ runs along the $x$-axis to demonstrate that the spin can sharpen over a time scale $\sim L^2$. This time scale characterizes the {spin}-sharp phase ($p > p_{\#}$). We used 30\,000 samples when $L = 8$ to $16$; and 10\,000 samples when $L = 18$.
    }
    \label{fig:CS_2}
\end{figure}

Figure \ref{fig:CS_2} reveals the phases' spin-sharpening time scales: $\sim L^2$ in the spin-sharp phase and $\sim L^3$ in the spin-fuzzy phase. A simple argument supports the latter~\citep{huse2023}: $\ket{\psi_{\rm i}}$ corresponds to an eigenvalue $s(s+1) \in \{1, 2\}$ of 
$\vec{S}^2 = \sum_{j,k} \vec{\sigma}_j \cdot \vec{\sigma}_k$. 
The system contains $ \sim L^2$ pairs $(j, k)$. One might expect all pairs to contribute roughly equally to $\langle \vec{S}^2 \rangle$, by ergodicity, in the spin-fuzzy phase. Hence
$\langle \vec{\sigma}_j \cdot \vec{\sigma}_k \rangle \sim s(s+1) / L^2$.
To identify $s(s+1)$, we therefore must measure $L^2$ correlators $\langle \vec{\sigma}_j \cdot \vec{\sigma}_k \rangle$. Measuring one correlator with an imprecision $\sim 1/L$ requires $\sim L^2$ measurements. We hence need $\sim L^4$ measurements total. Since (const.)$L$ measurements occur per time step, the spin should sharpen in a time $\sim L^3$.

{Our identification} of a spin-sharpening transition at $p_\#$ is subject to at least two caveats. First, the crossing point drifts to larger $p$ as $L$ increases (perhaps coalescing with the purification transition at $p_c$ as $L \to \infty$). Second, the scaling ansatz we chose for the data collapse in Fig.~\ref{fig:CS}b may not be valid. The ansatz implies that the time scale for a size-$L$ system to sharpen increases more quickly than $L^2$ for $p < p_\#$ and more slowly than $L^2$ for $p > p_\#$. However, our data for $p > p_\#$ (see Fig.~\ref{fig:spin_sharpening} in Appendix \ref{sec:AddNum}) is compatible with a sharpening time scale $\sim L^2$ deep in the critical phase. If the sharpening time indeed scales as $L^2$ throughout the critical phase, the crossing in Fig.~\ref{fig:CS}a must be a finite-size artifact. Precisely identifying $p_\#$ and the sharpening transition's nature is outside the scope of this work, due to the paucity of $L$ values accessible in exact computations. We defer a detailed analysis of the spin-sharpening time scales to future work.

Finally, the spin-sharpening transition suggests a postselection-free means of observing a measurement-induced transition experimentally~\citep{noel2022measurement,barratt2022transitions,hoke2023quantum, koh2022experimental}: identify whether an observer can learn $s$ from measurement outcomes in a given time interval. This learning would require ``decoders'' for estimating $s$ from the outcomes. The decoders' accuracy, as a function of the measurement rate, would need to be tested. In principle, one can learn $s$ most accurately via brute-force decoding~\cite{barratt2022transitions}. One would, upon running the circuit and obtaining the measurement outcomes, simulate the circuit, postselected on the observed outcomes and operating on a state in the $s_0$ sector. Next, the simulation {would be repeated} with an initial state in the $s_1$ sector. From each simulation, {the} probability that $s_0$ (or $s_1$) had engendered the observed outcomes {could be inferred}.

However, this approach generically costs exponential-in-$L$ time (even if a quantum computer performs the simulation, due to the postselection). Special classes of monitored dynamics~\citep{noel2022measurement,barratt2022transitions,hoke2023quantum, koh2022experimental} may allow for approximate decoders that can be implemented efficiently on classical or quantum computers without postselection. In this case, the transition's nature will depend on both the circuit and the decoder and may differ, in location or universality class, from the spin-sharpening transition observed under optimal decoding. We leave for future work the problem of designing efficient decoders for spin-sharpening transitions.

\section{Effective-Hamiltonian description of the monitored dynamics}\label{Sec:StaMecMod}

To complement the numerics, we derive an effective statistical-mechanics model: a description of the monitored evolution as imaginary-time evolution under an effective Hamiltonian acting on copies (replicas) of the system. In the rest of this section, we describe the model. We elucidate its ground and low-lying excited states in Sec.~\ref{sec_ground_low_p}. Leveraging these results, we elucidate the monitored circuit's purification transition in Sec.~\ref{sec_eff_H_purification}. Section~\ref{sec_eff_H_area_law} explains the circuit's lack of an area law.

In the statistical-mechanics model, measurement outcomes act as quenched disorder for a quantum trajectory. A replica trick is needed to average nonlinear quantities, such as entanglement, over trajectories~\cite{jian2020measurement, bao2020theory}. One must average $Q$ replicas of the density matrix, $\rho^{\otimes Q}$. In the replica limit, $Q\rightarrow 1$. More precisely, we want to calculate
\begin{align}
\overline{\rho^{(Q)}(t)} 
&= \sum_{\vec{m}} \int dU \left(K_{\vec{m},U}^{\vphantom\dagger}\rho_0K_{\vec{m},U}^\dagger \right)^{\otimes Q}
\, ,
\end{align}
dependent on the evolution operator
$K_{\vec{m},U} \equiv\prod_{\ell=1}^{2t} P_{\ell,\vec{m}}U_\ell$. $ U_\ell $ denotes the unitary implemented by circuit layer $\ell$. $P_{\ell,\vec{m}}$ denotes the projector onto the subspace associated with the list $\vec{m}$ of outcomes yielded by the measurements at time step $i$. $\int dU$ denotes an average over the 
SU(2)-symmetric gates (with the appropriate probability measure). 

An alteration to the circuit model will facilitate the analytics: we deform the discrete-time, strong-measurement circuit dynamics into a continuous-time version. We replace the gates with Hamiltonian evolutions over infinitesimal time steps, and infinitesimally weak measurements replace the projective measurements. We expect the continuous-time deformation to preserve the purification/entanglement and charge-sharpening transitions' universal scaling properties. The reasons are analogous examples~\cite{barratt2022transitions} and the system's lack of time-translation symmetry.
 
Here, we summarize the resulting Hamiltonian description. Appendix~\ref{appendix: eff H} contains a detailed derivation. The effective Hamiltonian equals a sum of contributions from the unitary dynamics and the weak measurements: 
$H^\mathrm{eff} = H^{\mathrm{u}} + H^{\rm m}$. 
The terms are 
\begin{align}
    H^{\rm u} &= -J \sum_{i} \left[ \sum_{a=1}^Q\left( \vec{S}^a_{i}\cdot\vec{S}^a_{i+1} - \vec{S}^{a*}_{i}\cdot\vec{S}^{a*}_{i+1} \right)\right]^2
    \; \text{and} \label{eq: H_u}\\
    H^{\rm m}  &=
    \gamma \sum_{i} \sum_{a,b=1}^{2Q} \left( \vec{S}^a_{i}\cdot\vec{S}^a_{i+1} \right) 
    \Pi_{a,b}  
    \left( \vec{S}^b_{i}\cdot\vec{S}^b_{i+1} \right) . \label{eq: H_m}
\end{align}
The coupling constant $J$ encapsulates the unitary dynamics' scrambling power; and $\gamma$, the weak measurements' strength.
$P_{i,j}^a$ is the projector onto the singlet sector of spins $i$ and $j$ in replica copy $a$. Equations~\eqref{eq: H_u}--\eqref{eq: H_m} have two kinds of summations over the replica index, $a$. When $a$ runs from 1 to $Q$, the summation is over forward copies of the replicas. $a^*$ represents the corresponding backward copy. When $a$ runs from 1 to $2Q$, the summation is over both the backward and forward copies. The projector $\Pi_{a,b} = \delta_{ab}-\frac{1}{2Q}$ is onto inter-replica fluctuation modes. The associated term in Eq.~\eqref{eq: H_m} is minimized when $\vec{S}_i\cdot\vec{S}_{i+1}$ yields the same value, operating on any replica, as operating on any other. If the measurements are projective, all the replicas must yield the same measurement outcome. If the measurements are weak, as above, this restriction is softened; a finite energy cost accompanies inter-replica fluctuations in the measured operator $\vec{S}\cdot\vec{S}$.

The effective Hamiltonian has a left/right $S_Q\times S_Q$ symmetry: $H^\mathrm{eff}$ remains invariant under permutations of the $Q$ forward copies and permutations of the $Q$ backward copies. The monitored dynamics map to imaginary-time evolution under $H^\mathrm{eff}$ (in the replica limit $Q \to 1$). Thus, we must understand this Hamiltonian's low-energy properties to understand the monitored dynamics' late-time properties.  

\subsection{Ground state and collective excitations at low measurement rates}
\label{sec_ground_low_p}

We begin with a measurement-free model: $ \gamma=0 $ in Eq.~\eqref{eq: H_m}. A ground state is a configuration that, when acted on by $\sum_{a=1}^Q\left( \vec{S}^a_{i}\cdot\vec{S}^a_{i+1} - \vec{S}^{a*}_{i}\cdot\vec{S}^{a*}_{i+1} \right)$ for any nearest-neighbor pair $(i, i+1)$, vanishes. Such a configuration is achievable if and only if, for some pairing of $(a,b^*)$, $\vec{S}^a_{i}\cdot\vec{S}^a_{i+1} = \vec{S}^{b*}_{i}\cdot\vec{S}^{b*}_{i+1}$ for all $i$. The ground states thus can be labeled by all such pairings $(a,b^*)$. Furthermore, the ground states are represented by the elements $\sigma$ of the permutation group $S_Q$ such that $\vec{S}^a_{i}\cdot\vec{S}^a_{i+1} = \vec{S}^{\sigma(a)*}_{i}\cdot\vec{S}^{\sigma(a)*}_{i+1}$. To satisfy this condition for all $i$, the interaction must be ferromagnetic, precluding frustration. We show in Appendix~\ref{appendix: eff H} that the ground space of  $H^{\rm u}$ is that of an SU(4) ferromagnet. The ground states can thus be labeled as $\superket{\otimes_{i=1}^L \sigma}$. The permutation $\sigma\in S_Q$, and the tensor product emphasizes the pairings' uniformity across space.\footnote{As noted, the ground space has a degeneracy labeled by the ground states of the SU(4) ferromagnet. The label depends on the initial state. We drop the label from our notation for simplicity, as the label does not impact the following discussion.} Importantly, the ground space breaks the discrete symmetry $S_Q\times S_Q$ to $S_Q$.

We now briefly sketch the low-lying energy eigenstates.\footnote{The gaps between these eigenstates' energies and the ground-state energy vanishes in the thermodynamic limit but remains nonzero at finite $L$.} If $\gamma=0$, the excitations over a symmetry-broken state $\superket{\otimes_{i=1}^L \sigma}$ are described by $Q$ decoupled SU(4) ferromagnetic chains, each formed from two SU(2) chains. Let us focus on one SU(4) chain. An SU(4) ferromagnet's Goldstone modes live on a 6-dimensional manifold. They result in three gapless modes with energies vanishing as $L^{-z}$, wherein $z=2$. These gapless modes are of two types: Two modes arise from fluctuations within single SU(2) spin chains. The third mode arises from collective fluctuations of the two SU(2) chains. In summary, $Q$ replicas lead to $2Q$ diffusive ($z{=}2$) modes associated with fluctuations within single SU(2) chains, plus $Q$ diffusive modes associated with inter-replica fluctuations. As noted above, measurements affect only the inter-replica fluctuations and thus couple the $Q$ inter-replica modes [Eq.~\eqref{eq: H_m}]. 

Consider increasing the measurement parameter $\gamma$ from 0.
As in U(1)-symmetric circuits~\cite{barratt2022field}, measurements gap out some inter-replica degrees of freedom. Furthermore, the inter-replica gapless modes reduce to one diffusive mode (corresponding to the fluctuations in the replicas' average) and $Q-1$ relativistic ballistic ($z{=}1$) modes, which describe inter-replica fluctuations. These ballistic modes cause the R\'enyi entropies with indices $n>1$ to grow ballistically in the presence of measurements~\cite{agrawal2022entanglement,barratt2022field}. The diffusive inter-replica modes are well-defined only for symmetry-broken states whose forward and backward copies are paired explicitly. Thus, these modes are expected to survive only in the replica-symmetry-broken phase (volume-law phase). However, the $2Q$ intrachain-fluctuation SU(2) modes do not depend on such pairings. Hence these modes are expected to exist at all measurement strengths and so in the critical phase. As we discuss below, these surviving gapless $z{=}2$ modes likely underlie two circuit behaviors that we observed: the $L^2$ purification time scale and the absence of area-law entanglement.

\subsection{Purification}
\label{sec_eff_H_purification}

Using the formalism above, we can understand the purification of an initially maximally mixed state, $\superket{\otimes^L_{i=1} e} $. The $ e \in S_Q$ denotes a permutation that pairs replica $ a $ with $ a^* $. At a late time $t$, the density matrix's trajectory-averaged purity $\Pi(t)$ is given by
\begin{align}
	\Pi(t) = \lim_{Q\rightarrow 1} \frac{\superbra{\otimes^{L}_{i=1}g}e^{-\beta H^{\mathrm{eff}}}\superket{\otimes^{L}_{i=1} e}}{\superbra{\otimes^{L}_{i=1}e}e^{-\beta H^{\mathrm{eff}}}\superket{\otimes^{L}_{i=1} e}} \, .
\end{align}
$ g $ denotes the transposition that swaps replica 1 with 2* and 2 with 1* while acting as the identity on the other replicas~\cite{jian2020measurement,bao2020theory}.
The purification time is when $ \Pi(t) $ becomes $ \mathcal{O}(1) $. 
In the absence of measurements, $\gamma=0$ an initially maximally mixed state will fail to purify and will have $\Pi(t)=\frac{1}{2^L}$ for all times. Indeed, at $ \gamma = 0 $, $ \superket{\otimes^{L}_{i=1} e} $ is a ground state of $ H^\mathrm{eff} $ and has vanishing energy. Thus, $ \Pi(t) |_{\gamma=0} = \lim_{Q\to1}\llangle {\otimes^{L}_{i=1}g} | \otimes^{L}_{i=1} e \rrangle = 1/2^L$. Excitations of the discrete-symmetry-broken phase are gapped domain-wall configurations. Therefore, we expect this phase to be stable under the strengthening of the weak measurements to low rates $\gamma$. For the system to transition between different replica-symmetry-broken ground states, a domain wall must tunnel across the entire system. Such transitions are thus expected occur over an exponential-in-system-size time, which we identify as the purification time: $t_{\rm P} \sim {\rm e}^L$. {This exponential scaling is a well-established result in the absence of symmetry}~\cite{skinner2019measurement,li2018quantum, li2019measurement, jian2020measurement}{, and} the SU(2) symmetry has little bearing on the replica-symmetry breaking; essentially identical arguments apply in the symmetry's absence. This behavior contrasts with that of lattice magnets that have continuous symmetries. There, a symmetry-broken state can be deformed smoothly into another symmetry-broken state over a $ \mathrm{poly}(L) $  time scale. Interestingly, monitored free fermions have an emergent, continuous inter-replica symmetry (as opposed to our discrete $ S_Q $ symmetry), resulting in linear purification times~\cite{bao2021symmetry,poboiko2023theory,fava2023nonlinear,jian2023measurementinduced}.

{On the one hand, the domain walls correspond to the broken discrete permutation symmetry in the volume law phase. On the other hand, the gapless modes do not mix permutations. 
The discrete-symmetry-broken phase exists regardless of whether there are also continuous symmetries.}

The replica symmetry is restored at sufficiently high measurement strengths; the argument for $t_{\rm P} \sim {\rm e}^L$ breaks down. Instead, the purification time depends on the effective Hamiltonian's energy gap. We conjecture that this gap scales as $1/L^2$, due to the gapless modes associated with the previous subsection's $2Q$ $z{=}2$ modes. This gap scaling results in a purification time $t_{\rm{P}}\sim L^2$.

\subsection{Absence of area law under strong measurements}
\label{sec_eff_H_area_law}

We can establish the absence of an area-law phase at any measurement rate by adapting a Lieb-Shultz-Mattis-type-anomaly argument to the spin model in the replica trick with $2Q$ copies, as first argued in~\cite{nahum2020entanglement}. (See also~\cite{PhysRevX.8.031028, ma2022average}, which generalize this result to statistical symmetries.) Each of the $2Q$ copies has SU(2) symmetry, and the replica symmetry permutes the replicas. Additionally, under averaging over the measurements and circuit elements, the replica model has a $\mathbb{Z}$ lattice-translation symmetry. Overall, the statistical-mechanics model's symmetry group is  $G= \mathbb{Z} \times \left[ ( {\rm SU}(2)^{\times Q} \rtimes S_Q) \times ( {\rm SU}(2)^{\times Q} \rtimes S_Q) \rtimes \mathbb{Z}_2   \right]$~\cite{bao2021symmetry}. Each site contains one (projective) spin-$1/2$ representation of each replica factor of SU(2). Therefore, there is a mixed anomaly between translation symmetry and the SU(2) spin-rotation symmetry. This anomaly rules out the possibility of a featureless (short-range-entangled, symmetry-preserving) ground state. Moreover, na\"{i}vely applying the Mermin-Wagner theorem rules out spontaneous breaking of the SU(2) symmetry (although subtle examples may violate this principle in the replica limit~\cite{jacobsen2003dense}). Furthermore, we observe no tendency towards any spontaneous breaking of the lattice-translation symmetry. These arguments suggest that no area-law phase can arise, even in the measurement-only limit.

We can obtain further insight into the strong-measurement regime through our mapping to an effective-Hamiltonian model. Each replica has an SU(2) symmetry, which leads to gapless modes, as noted above, with ferromagnetic interactions and thus $z{=}2$ dynamics. These conclusions are consistent with the critical phase 
observed in our numerics. 

Using the formalism above, one can calculate R\'enyi entropies of the reduced density matrix of an interval A. Let $\rho$ denote any single-copy pure state; and $\superket{\rho}$, the $Q$-replica defined on $2Q$ copies of the Hilbert space. We focus on the R\'enyi index $n=2$: 
\begin{align}
    \label{eq_renyi_1}
    e^{-S^2(\rho_A)} = \lim_{Q\rightarrow 1} \frac{\superbra{\otimes_{i=1}^L g^{A}_i} e^{-\beta H^{\mathrm{eff}}} \superket{\rho}}{\superbra{\otimes_{i=1}^L e} e^{-\beta H^{\mathrm{eff}}} \superket{\rho}} \, .
\end{align}
$g^{A}_i=e$ if $i$ does not belong to the interval $A$, $g^A_i=(12)(34)...(2k{-}1\ 2k)$ 
if $i\in A$, and $Q=2k+1$. 
We define a ``twist'' permutation $\tau$ such that 
$\tau \left( \otimes_{i=1}^L g^{A}_i \right)  = \otimes_{i=1}^L e$. Using $\tau$, we can rewrite~\eqref{eq_renyi_1}.
Since the initial state is pure, 
$\tau \superket{\rho} = \superket{\rho}$, and
\begin{align}
    e^{-S^2(\rho_A)} = \lim_{Q\rightarrow 1} \frac{\superbra{\otimes_{i=1}^L e} \tau^{-1}e^{-\beta H^{\mathrm{eff}}} \tau \superket{\rho}} {\superbra{\otimes_{i=1}^L e} e^{-\beta H^{\mathrm{eff}}} \superket{\rho}} \, .
\end{align}
$\tau$, operating on $H^\mathrm{eff}$, introduces a twist operator at the interval's boundary, $\partial A$: $\tau^{-1}e^{-\beta {H^{\mathrm{eff}}}} \tau \equiv T_{\partial A}\ e^{-\beta H^{\mathrm{eff}}}$. Hence a size-$|A|$ interval's R\'enyi-2 entropy is related to the two-point correlator of the twist operator acting on sites separated by a distance $|A|$. In the infrequent-measurement phase with spontaneously broken replica symmetry, this correlator decays exponentially, leading to volume-law R\'enyi entropies. Intuitively, domain walls in this discrete ferromagnetic phase have finite line tensions. Hence creating a domain wall costs an extensive (volume-law) amount of free energy. Under frequent measurements, in the putative critical phase, the permutation degrees of freedom are gapped. The twist operator should {likely}, instead, couple to the remaining low-energy SU(2) modes. Analyzing the critical phase's nature, requiring the effective Hamiltonian's replica limit, presents a clear challenge for future work. {This critical phase, with (presumably) logarithmic entanglement scaling appears unrelated to the low-energy physics of quenched random Heisenberg spin chains, whose entanglement scales similarly}~\cite{PhysRevLett.93.260602}. {Quenched random Heisenberg chains have $z = \infty$, in contrast with $z=2$ that we observe numerically.}

\section{Outlook} \label{sec:OutLoo}

We studied the dynamics of monitored random circuits with SU(2) symmetry, i.e., with three noncommuting charges: the total spin angular momentum's components. First, we numerically discovered a purification transition between a mixed phase (at $p<p_{\rm c} \approx 0.35$) and a critical phase (at $p>p_{\rm c}$). In the critical phase, the purification time scales as $t_{\rm P}\sim L^2$. The purification transition doubles as an entanglement transition, which separates volume-law (at $p<p_{\rm c}$) and subextensive (logarithmic or small-power-law, at $p>p_{\rm c}$) entanglement scalings. Even in the measurement-only limit (at $p=1$), the symmetry's non-Abelian nature enables nontrivial entanglement scaling. Additionally, we observed a spin-sharpening transition across which there is a parametric change in the time at which one can (in principle) learn the system's total spin by monitoring measurements. The time scale is $t\sim L^2$ in the  ``spin-sharp'' phase and $t \sim L^3$ in the ``spin-fuzzy'' phase.

Finally, we interpreted our results within an effective replica statistical-mechanics model. The model supports the mixed-phase prediction that $t_{\rm P} \sim e^L$. Also, the model hints at a possible spin-wave mechanism for the $t_{\rm P} \sim L^2$ dynamics in the critical phase. Furthermore, a Lieb-Schultz-Mattis-type anomaly obstruction implies the absence of an area-law phase. Instead, the entanglement should scale logarithmically with $L$ in the critical phase, consistently with our numerics.

Our results open several opportunities for future work. One is to understand the purification/entanglement and sharpening transitions analytically. Second, one might leverage spin sharpening to observe an MIPT experimentally, avoiding the postselection problem (Sec.~\ref{sec:int}). The thermodynamics of noncommuting charges have already been observed experimentally with trapped ions~\cite{kranzl2022experimental}. Superconducting qubits, quantum dots, and spinful fermionic atoms are natural candidates, too~\cite{nyh2020noncommuting, nyh2022how}. 
Third, our system offers a playground for numerically exploring the recent result that non-Abelian symmetries constrain local unitary circuits more than Abelian symmetries do and so may constrain chaos more~\cite{marvian2022restrictions,marvian2021qudit,marvian2022rotationally,marvian2023non}.
Finally, efficient classical and quantum spin-sharpening decoders merit exploration.

\begin{acknowledgments}
We thank Ehud Altman, Fergus Barratt, David Huse, Andreas Ludwig, and Xiaoliang Qi for helpful discussions. We also thank Michael Gullans for galvanizing this collaboration. This work received support from the National Science Foundation under QLCI grant OMA-2120757 (N.Y.H.), the John Templeton Foundation under award No. 62422 (N.Y.H. and S.M.), the Air Force Office of Scientific Research under Grant No. FA9550-21-1-0123 (R.V.), the Alfred P. Sloan Foundation through Sloan Research Fellowships (A.C.P. and R.V.), National Science Foundation under NSF Grants No. DMR-1653271 (S.G.), and the Vanier C.G.S. (S.M.). This work was supported by the Simons Collaboration on Ultra-Quantum Matter, which is a grant from the Simons Foundation (651440, U.A.)

\end{acknowledgments}

\begin{appendices}

\renewcommand{\thesection}{\Alph{section}}
\renewcommand{\thesubsection}{\Alph{section} \arabic{subsection}}
\renewcommand{\thesubsubsection}{\Alph{section} \arabic{subsection} \roman{subsubsection}}
\makeatletter\@addtoreset{equation}{section}
\def\theequation{\thesection\arabic{equation}}

\section{Additional numerics elucidating the entanglement dynamics and spin sharpening} \label{sec:AddNum}

\begin{figure}
    \centering
    \includegraphics[width=\columnwidth]{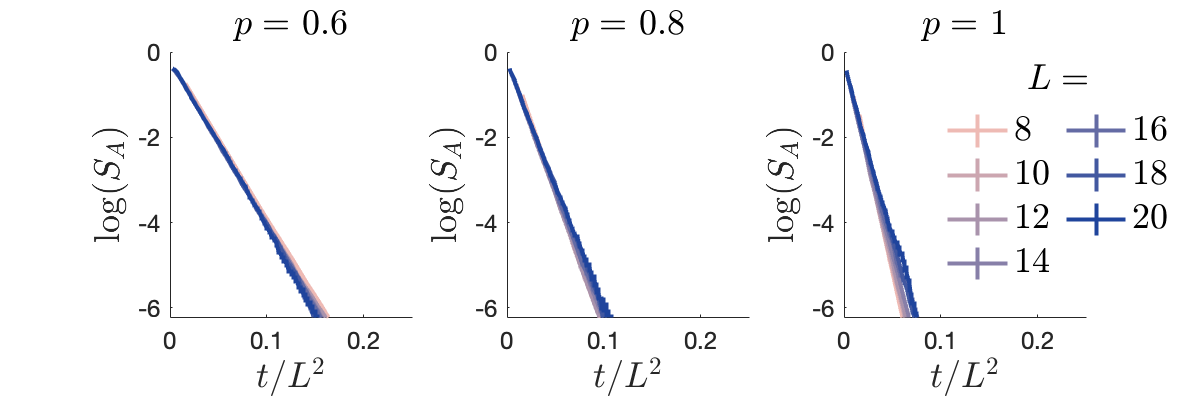}
    \caption{\caphead{{The purification time still reveals a $z=2$ phase for $p > 0.4$.}} {The entropy $S_A$ quantifies the ancilla qubit's entanglement with the system. We plot $\log(S_A)$ for clarity, as $S_A$ decays exponentially. $t/L^2$ runs along the $x$-axis to demonstrate the existence of a phase in which the system purifies over a time scale $t_{\rm P} \sim L^2$. We used 30\,000 samples when $L = 8$ to $16$; 10\,000 samples when $L = 18$; and 1\,500 samples when $L=20$. The $y$-axis's lower limit is $\log(10^{-3}) \approx -6.91$.} }
    \label{fig:remaining_purification}
\end{figure}

{In Fig.}~\ref{fig:purification_times}{, we plot $\log(S_A)$ [see Eq.}~\eqref{eq:EE_onestate}{] against $t/L^2$. We claimed the different-$L$ curves collapse for $p>p_c \approx 0.35$ and presented the plots for up to $p = 0.4$. To confirm that the curves remain collapsed for larger $p$ we plot $p = 0.6, 0.8,$ and $1.0$ in Fig.}~\ref{fig:remaining_purification}.

\begin{figure}
    \centering
    \includegraphics[width=0.49\columnwidth]{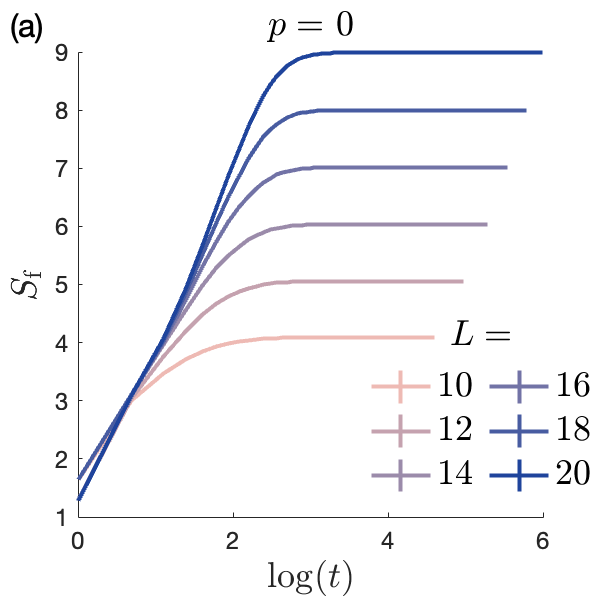}
    \includegraphics[width=0.49\columnwidth]{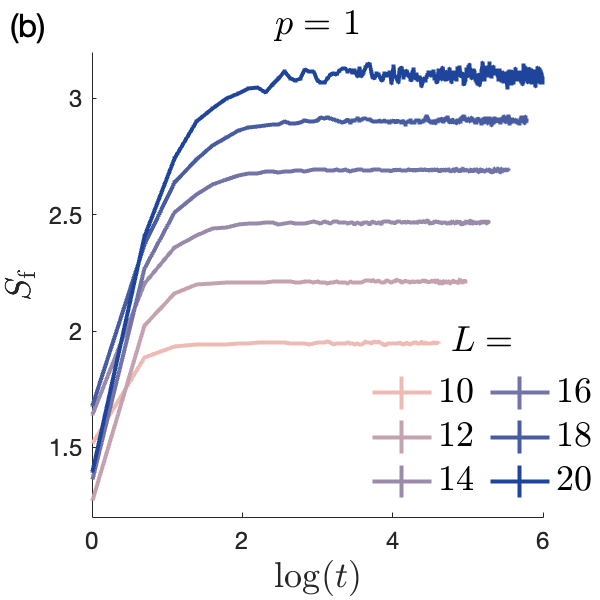}
    \caption{\caphead{The bipartite entanglement entropy saturates after $L^2$ time steps.} At the extreme $p$ values $p=0,1$, $S_{\rm f}$ quits changing (to within minor fluctuations).}\label{fig:EE_vs_time}
\end{figure}

In Sec.~\ref{sec:EntDyn}, we claimed that $L^2$ time steps suffice for the bipartite entanglement entropy, $S_{\rm f}$, to plateau. Figure~\ref{fig:EE_vs_time} justifies this claim, presenting $S_{\rm f}$ as a function of $\log(t)$ for $\leq L^2$ time steps at the extreme values $p = 0, 1$. At both extrema, $S_{\rm f}$ stops changing (to within minor fluctuations) by $L^2$ time steps.

\begin{figure}
    \centering
     \includegraphics[width=\columnwidth]{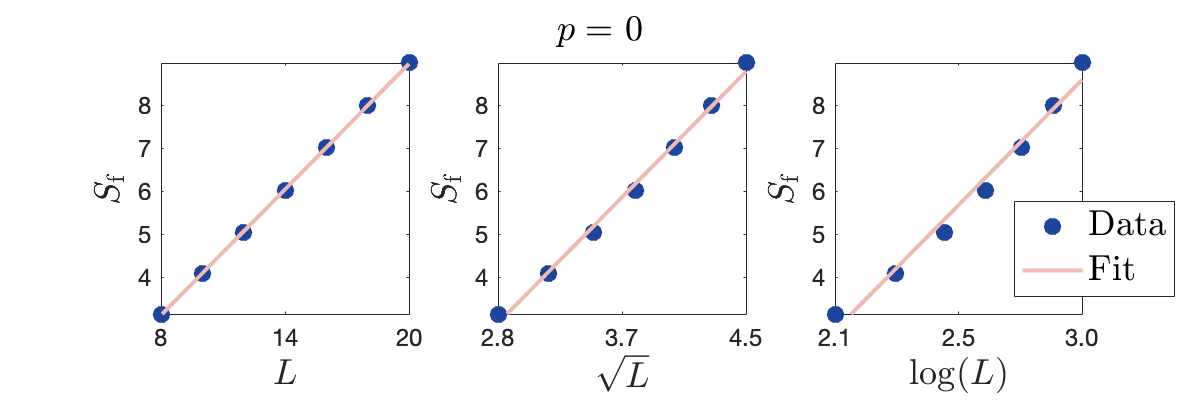}\\
     \includegraphics[width=\columnwidth]{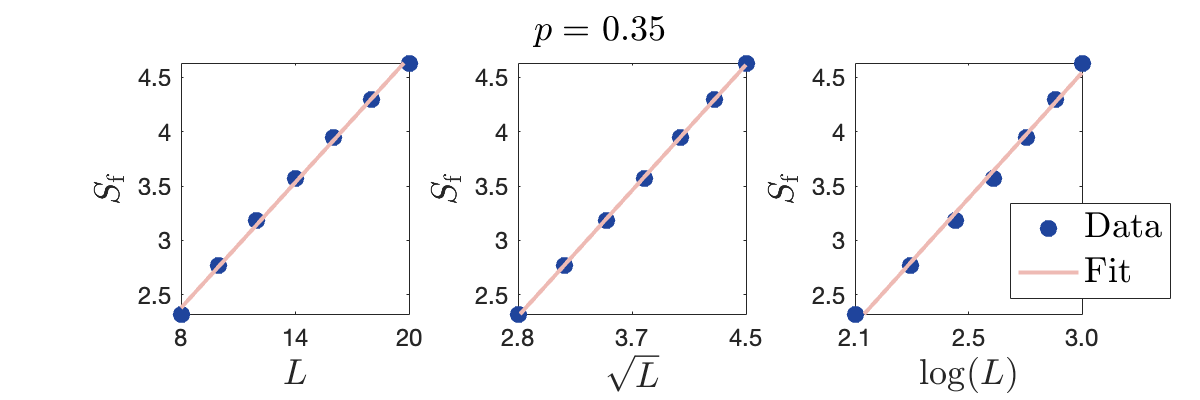}\\
     \includegraphics[width=\columnwidth]{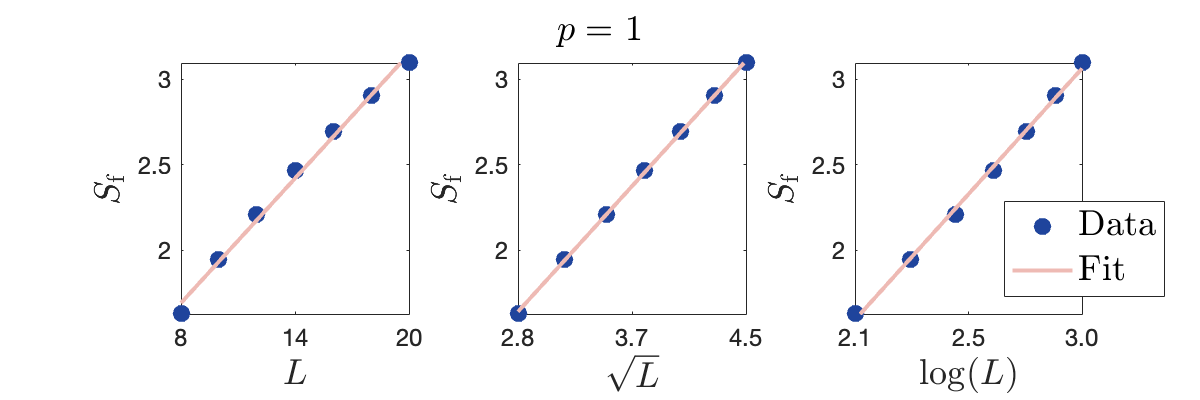}
    \caption{\caphead{Long-time bipartite entanglement entropy vs. system size.} At $p=0$, $S_{\rm f} \sim L$, signaling a volume law. At $p=1$, the entropy scales logarithmically or as a small power law:
    $S_{\rm f} \sim \log(L)$, or $S_{\rm f} \sim \sqrt{L}$.}
    \label{fig:example_fits}
\end{figure}

Section~\ref{sec:EntDyn} also discussed different fittings for $S_{\rm f}$ versus $L$. Figure~\ref{fig:example_fits} presents three fittings [$L$, $\log(L)$, and $\sqrt{L}$] at each of three measurement rates ($p =0$, $p=1$, and $p \approx p_{\rm c}$). At $p=0$, the linear fit is the best. This observation is consistent with the existence of a volume-law phase at $p=0$. At $p = 0.35 \approx p_{\rm c}$, it is unclear which fit is most accurate. However, the two nonlinear fits are visibly best. The $p=1$ fits resemble the $p = 0.35$ ones.

\begin{figure*}
    \centering
    \includegraphics[width=2\columnwidth]{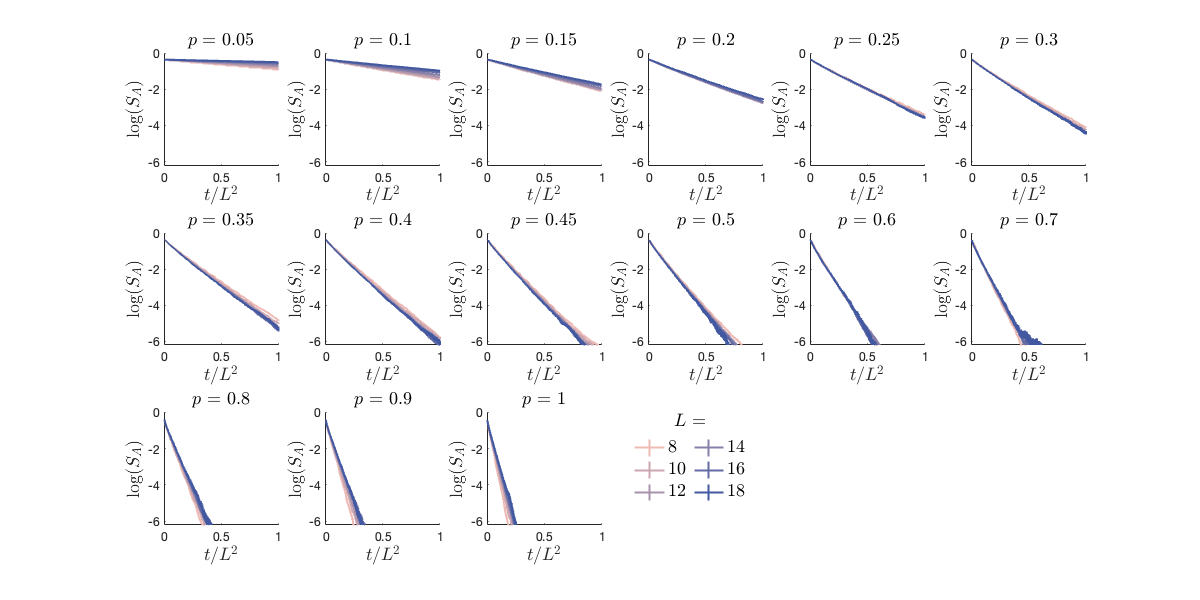}
    \caption{\caphead{In the critical phase, the numerics are consistent with a $\sim L^2$ sharpening time scale.}
    The entropy $S_A$ quantifies the ancilla qubit's entanglement with the system. We plot $\log(S_A)$ for clarity, as $S_A$ decays exponentially. $t/L^2$ runs along the $x$-axis to demonstrate the numerics are consistent with a $\sim L^2$ sharpening time scale. We used 30\,000 samples when $L = 8$ to $14$; and 10\,000 samples when $L = 16$ to $L = 18$. The $y$-axis's lower limit is $\log(10^{-3}) \approx -6.91$.
    }
    \label{fig:spin_sharpening}
\end{figure*}

Section \ref{Sec:ChaSha} claimed that our $p > p_{\#}$ data are compatible with a sharpening time scale $\sim L^2$ deep in the critical phase. Figure \ref{fig:spin_sharpening} justifies this claim. We plot $\log(S_A)$ against $t/L^2$ at various $p$ values. The initial collapse occurs at $p > p_{\#}$. The $L{=}8$ numerics deviate from the collapse when $p \in [0.35, 0.45] \cup [0.8, 1]$. We suspect that these deviations arise from finite-size effects.

\section{Mutual information}\label{sec:Cor}

In Sec.~\ref{sec:Num}, we studied purification and entanglement dynamics. We complement that numerical analysis by studying mutual information. To introduce the mutual information, we consider a quantum system in a state $\ket{\psi}$. Let $\mathcal{A}$ and $\mathcal{B}$ denote subsystems. The reduced state of $\mathcal{A}$ is $\rho_{\mathcal{A}} \coloneqq \tr_{\bar{\mathcal{A}}}(\dyad{\psi})$. The reduced states of $\mathcal{B}$ and $\mathcal{A}\mathcal{B}$ are defined analogously. The mutual information between $\mathcal{A}$ and $\mathcal{B}$ is
\begin{equation}
    I(\mathcal{A}:\mathcal{B}) \coloneqq S(\rho_{\mathcal{A}}) + S(\rho_{\mathcal{B}}) - S(\rho_{\mathcal{AB}}).
\end{equation}
The mutual information upper-bounds equal-time correlators between local operators acting nontrivially on $A$ alone and on $B$ alone~\cite{wolf2008area}. We denote by $I^{(1)}_{j,k}$ the mutual information between sites $j$ and $k$. We denote by $I^{(2)}_{j,k}$ the mutual information between the pair $(j, j+1)$ and the pair $(k, k+1)$.

\begin{figure}
     \centering
     \includegraphics[width=0.49\columnwidth]{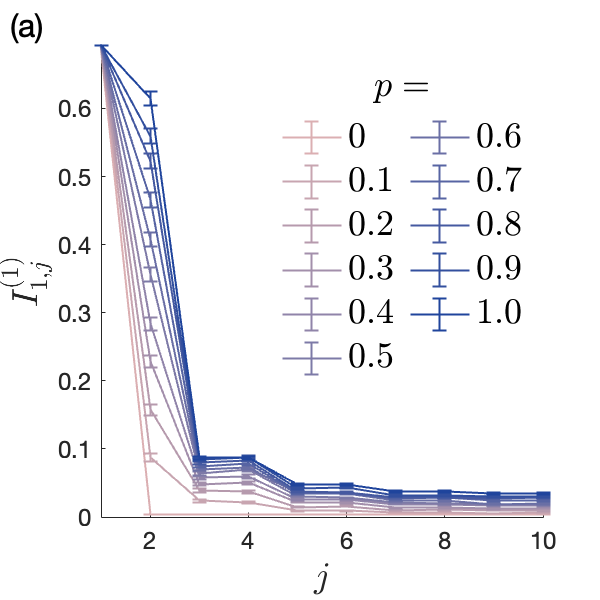}
     \includegraphics[width=0.49\columnwidth]{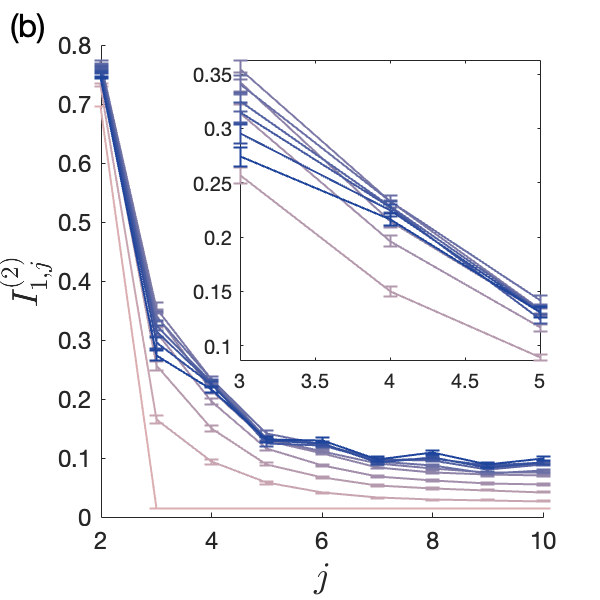}
    \caption{\caphead{Mutual information between sites.}
    The mutual information between (a) sites $1$ and $j$ decays more quickly than between (b) sites $(1,2)$ and $(j,j+1)$. The inset highlights how $I^{(2)}_{j,k}$ increases and then decreases as $p$ grows, for some $j$. The errorbars represent one standard deviation.
    }
    \label{fig:MI_distance}
\end{figure}

Figure~\ref{fig:MI_distance} presents $I^{(1)}_{1,j}$ and $I^{(2)}_{1,j}$, plotted against $j$, at $L=20$. $I^{(1)}_{1,j}$ grows with $p$ and rapidly decays with $j$.\footnote{Throughout these numerics, the last layer of gates was applied on the odd bonds (sites 1 and 2, sites 3 and 4, etc.), leading to larger $I^{(1)}_{j,j+1}$ for odd $j$ than even $j$.}
At all $p$, $I^{(1)}_{j,k}$ decays rapidly over distances $|j-k|$ larger than a few sites. 
This result is intuitive, since $I^{(1)}_{j,k}$ contains information about correlations between individual spin components, whereas the measurements and unitaries correlate spin-fusion channels (a property of two or more spins).

\begin{figure}
    \centering
     \includegraphics[width=0.49\columnwidth]{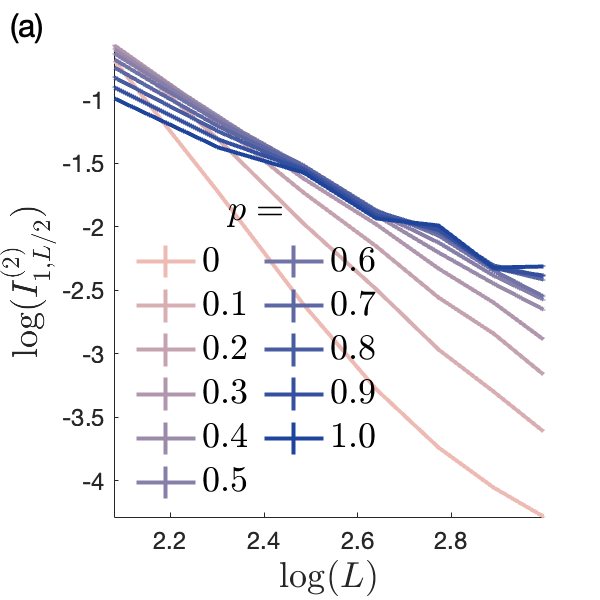}
     \includegraphics[width=0.49\columnwidth]{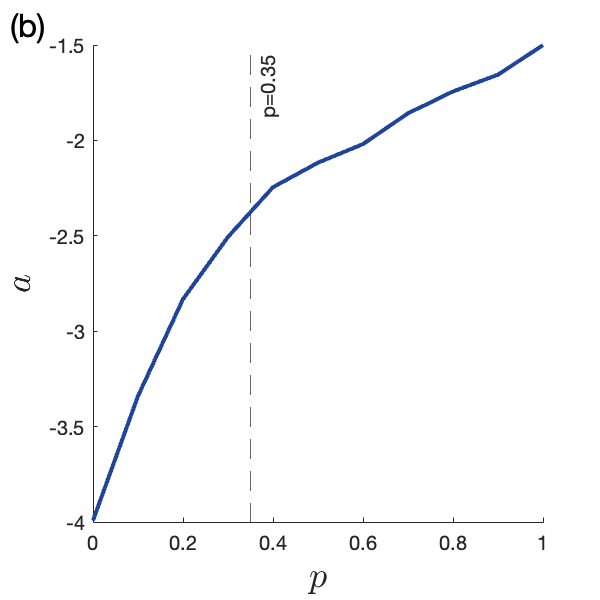}
    \caption{\caphead{Mutual information at antipodal sites.} 
    We call sites $1$ and $L/2$ antipodal. 
    (a) $\log(I^{(2)}_{1,L/2})$ is plotted against $\log(L)$ at several $L$ values. Using the fit function $\log(I^{(2)}_{1,L/2}) = a \log(L) + b$, we identify the critical exponent $a$ in $I^{(2)}_{1,L/2} \sim L^{a}$. 
    (b) Plotting $a$ against $p$, we find that $I^{(2)}_{1,L/2}$ decays as a power law in both phases, where $a$ seems to be drifting. 
    }
    \label{fig:antipodal_2MI}
\end{figure}

In comparison, $I^{(2)}_{j,k}$ decays more gradually with the distance $|j-k|$ at all $p>0$. For particular sites $j$ and $k$, $I^{(2)}_{j,k}$ may depend on $p$ nonmonotonically (Fig.~\ref{fig:MI_distance}). However, the asymptotic decay rate monotonically decreases as $p$ decreases. To explore this decay rate, we examine the mutual information between antipodal pairs of sites: $I^{(2)}_{1,L/2}$ (Fig.~\ref{fig:antipodal_2MI}a). Given the limitations on system size, we cannot convincingly determine the asymptotic decay's functional form. A power-law decay fits the data reasonably well (Fig.~\ref{fig:antipodal_2MI}b). The fitted power $a$ gradually decreases with $p$. Furthermore, $a$ changes qualitatively around $p_{\rm c} = 0.35$---from changing quickly with $p$, at $p<p_{\rm c}$, to drifting slowly near $-2$, at $p > p_{\rm c}$. Given the small range of system sizes available, exponential decay fits the data reasonably well, too; we cannot rule out this behavior. Yet, given the other critical scaling behavior at $p>p_{\rm c}$, we expect that power-law decay to be more natural in this regime. The data also prohibit confident distinction between (i) one power at $p>p_{\rm c}$, with drifts in the fitted exponent, due to finite-size corrections, and (ii) continuously evolving power laws (as would arise in, say, a Luttinger liquid).

\section{Effective Hamiltonian} \label{appendix: eff H}

In this appendix, we map the monitored dynamics onto the imaginary-time evolution of a replica effective Hamiltonian. 
Since the two-site gates are sampled independently, the above average over $U$ factorizes over averages of two-site SU(2)-symmetric gates. We parameterize the SU(2)-symmetric gates as \begin{align}
	U_{i,j} = P_t + e^{i\theta} P_s = e^{i\theta P_s} .
\end{align}
$P_t$ and $P_s$ denote projectors onto the triplet and singlet sectors of the SU(2) symmetry on two qubits. $ i$ and $j $ denote the qubits being acted on. We will suppress the subscript $i$ and $j$ unless they are needed to avoid confusion. This is similar Eq.~\eqref{eq:SU2unitary}, modulo a global phase factor. The variable $\theta$ is sampled from a random distribution---for example, from a uniform distribution between $0$ and $2\pi$. The $Q$ moment of the unitary gates is 
\begin{align}
	{\cal U} \equiv U^{\otimes Q} \otimes \left(U^{\dagger}\right)^{\otimes Q} = \mathrm{e}^{i\sum_a \theta \left( P^{(a)}_s - P^{(a^*)}_s \right)} \, .
	\end{align}
$ P_s^{(a)} $ denotes the projector onto the singlet sector for replica index $a$. The $ * $ symbol implies that the operator operates on the conjugate (backward) copy of the $ a $ replica (with $a=1, 2, \dots , Q$). We assume that $\theta$ is sampled from a Gaussian distribution $ P(\theta)  = \frac{1}{\sqrt{2\pi J}}e^{-\theta^2/(2J)} $, where $J$ is a large constant controlling 
the unitary dynamics' scrambling strength. Performing average over $ \theta $ yields \begin{align}
	\int d\theta \, P(\theta) \, {\cal U}   &= \mathrm{exp}\left( -J \left\{ \sum_{a=1}^Q 
 \left[ P^{(a)}_s - P^{(a^*)}_s \right] \right\}^2\right)  \\ 
	&\equiv \mathrm{exp}(-H^{\rm u}_{ij}). \nonumber
\end{align}

To model measurements in the continuum-time limit, we consider the weak-measurement protocol of~\cite{barratt2022field}. The action of the measurement of  a local operator $ O $ on the replica density matrix is described as \begin{align}
	&\sum_m P_{m}^{\otimes Q} \rho^{(Q)} P_{m}^{\otimes Q} \\
    & \rightarrow  
 \int dm \, e^{-\gamma \sum_{a=1}^Q \left[ (O^a - m)^2 + (O^{a^*}-m)^2 \right]} \rho^{(Q)}\nonumber \\
	= &\  \mathrm{exp}\left(-\gamma \sum_{a,b=1}^{2Q} O^a \Pi_{ab} O^b \right)  \rho^{(Q)} \\
	\equiv  &\ \mathrm{exp}(-\gamma H^{\rm m}) \rho^{(Q)}. \nonumber
\end{align}
$m$ denotes the weak measurement outcome, and $\Pi_{a,b} = \delta_{a,b} - 1/(2Q)$. We have identified $(a^*)$ with index $(Q+a)$ and $a=1,2,\dots,Q$. From now, on this identification will be implicit whenever the replica index $a$ is summed from $1$ to $ 2Q $. As in the main text, $\gamma$ denotes the weak-measurement strength. For SU(2)-symmetric systems, we measure the operators $O=\vec{S}_{i}\cdot\vec{S}_j$.

The expressions above are for averages of single 2-site unitary gates and measurements. We combine these local averages and assume a Trotter decomposition. The monitored evolution of the density matrix's averaged $Q^{\rm th}$ moment is given by imaginary-time evolution under an effective Hamiltonian $H^{\mathrm{eff}}$: $\overline{\rho^{(Q)}(t)} = e^{-t H^\mathrm{eff}} \rho_0^{(Q)}$.
The effective Hamiltonian decomposes as $H^\mathrm{eff} = \sum_{i} \left( H^{\rm u}_{i,i+1} +  H^{\rm m}_{i,i+1} \right)$, with\begin{align}
     H_{i,j}^{\rm m}   = \gamma \sum_{a,b} 
     \left( \vec{S}^a_{i} \cdot \vec{S}^a_j \right) \Pi_{a,b}  
     \left( \vec{S}^b_{i}\cdot\vec{S}^b_j \right) .
\end{align}
The long-time properties of $\overline{\rho^{(Q)}(t)}$ are thus described by low-temperature/ground-state properties of $H^\mathrm{eff}$. The Hamiltonian has a $S_Q \times S_Q$ symmetry, corresponding to the global permutation among the $Q$ forward replicas and $Q$ backward replicas.

A more illuminating way of understanding the structure of the Hamiltonian's ground states is to combine  spin-$1/2$ particles at replica $ a $ and $ \sigma(a) $, to form a fundamental representation of SU(4). The Hamiltonian's unitary part, in this identification, can be written as [the label $ \sigma $ signifies that we have combined replicas $ \LParen a,\sigma(a) \RParen $ to form an SU(4) representation] $H^{\rm u} = \frac{J}{2} \left(H_0[\sigma] + V^{\rm u}[\sigma]\right)$.
The $H_0[\sigma]$ are $Q$ copies of the SU(4) ferromagnet, and\begin{align}
	V^{\rm u}[\sigma] =& \sum_i\sum_{a<b} \left( \sw^a_{i,i+1}  - \sw_{i,i+1}^{\sigma(a)*} \right)\left(  \sw^b_{i,i+1}  - \sw_{i,i+1}^{\sigma(b)^*} \right). \label{eq: Unitary pert appendix}
\end{align}
$ \sw_{i,j}^a $ denotes the SWAP operator between the spins at sites $ i$ and $j $ in replica index $ a $: $ \sw_{i,j} = 1/2 + 2\vec{S}_i\cdot\vec{S}_j $. In terms of these SWAP operators, the SU(4) ferromagnet is
\begin{align}
    H_0[\sigma] =& \sum_{a=1}^Q \sum_i
    \left[ 1-\Big(\sw^a_{i,i+1} \Big)\left(\sw_{i,i+1}^{\sigma(a)*} \right)\right]. \label{eq: SU(4) ferro appendix}
\end{align} 

The measurement part of Hamiltonian also decomposes into two terms. One part is the SU(4) ferromagnet, $H^{\rm m} = \frac{\gamma}{Q} H_0[\sigma] - \frac{\gamma}{2Q} V^{\rm m}[\sigma]$,
where \begin{align}
	V^{\rm m}[\sigma] = \sum_i \sum_{a\neq b } \left( \sw^a_{i,i+1}+\sw^{\sigma(a)*}_{i,i+1} \right) \left( \sw^b_{i,i+1}+\sw^{\sigma(b)^*}_{i,i+1} \right). \label{eq: Meas pert appendix}
\end{align}

\end{appendices}

\onecolumngrid
\bibliography{apssamp}

\end{document}